# Interfacial Noncollinear Filtering of Spin Hall Currents

Dan-Yang Han,[1,2,3,*] Zi-An Wang,[1,2,*] Hong-Liang Chen,[4] Bo Li,[5] Wen-Jian Lu,[1] Yu-Ping Sun,[1,6,7] Liang Liu,[4,†] Shu-Hui Zhang,[3,‡] and Ding-Fu Shao[1,§]

[1] *Key Laboratory of Materials Physics, Institute of Solid State Physics, HFIPS, Chinese Academy of Sciences, Hefei 230031, China*
[2] *University of Science and Technology of China, Hefei 230026, China*
[3] *School of Physics, Nankai University, Tianjin 300071, China*
[4] *Tsung-Dao Lee Institute, Key Laboratory of Artificial Structures and Quantum Control (Ministry of Education), School of Physics and Astronomy, Shanghai Jiao Tong University, Shanghai 200240, China*
[5] *MOE Key Laboratory for Nonequilibrium Synthesis and Modulation of Condensed Matter, Shaanxi Province Key Laboratory of Quantum Information and Quantum Optoelectronic Devices, School of Physics, Xi'an Jiaotong University, Xi'an 710049, China*
[6] *Anhui Province Key Laboratory of Low-Energy Quantum Materials and Devices, High Magnetic Field Laboratory, HFIPS, Chinese Academy of Sciences, Hefei, Anhui 230031, China*
[7] *Collaborative Innovation Center of Microstructures, Nanjing University, Nanjing 210093, China*
[*] These authors contributed equally to this work.
[†] liul21@sjtu.edu.cn; [‡] shuhuizhang@nankai.edu.cn; [§] dfshao@issp.ac.cn

Spin Hall currents generated in nonmagnetic materials are conventionally regarded as bulk responses whose polarization is fixed by crystal symmetry. This view has motivated the search for intrinsically low-symmetry spin sources when unconventional spin polarizations are required. Here we point out that, in realistic heterostructures, the device-relevant quantity is not the fully symmetry-averaged bulk spin Hall current, but the emitted spin current transmitted across the interface. We therefore establish emitted spin currents as bulk–interface hybrid responses and propose interfacial noncollinear filtering as a mechanism to bypass the bulk-symmetry constraint. A low-symmetry interfacial spin-orbit field, generally noncollinear with the momentum-resolved spin polarization of the incident spin Hall current, imposes spin-dependent transmission and converts hidden momentum-resolved spin-polarization components into an observable unconventional emitted spin current. Using both a rotationally symmetric minimal model and a realistic high-symmetry Dirac-semimetal model, we show that conventional spin Hall sources can emit sizable out-of-plane spin currents when their hidden bulk spin Hall textures are selectively transmitted by the interfacial spin-orbit field. Our results reveal that spin-current polarization emerges from the cooperative action of bulk and interfacial responses, providing a strategy for reprogramming spin-current polarization in high-efficiency, CMOS-compatible spin Hall materials without relying on intrinsically low-symmetry bulk crystals or external symmetry-breaking schemes.

Spin currents constitute a central physical quantity in spintronics because their generation, propagation, and transfer enable information writing and readout in spin-based devices [1]. Among different generation mechanisms, the spin Hall effect in nonmagnetic materials is particularly important: it converts a charge current into a transverse pure spin current and allows heavy metals and semimetals to serve as efficient spin sources [2–5].

A key objective is therefore not only to achieve large spin Hall current magnitudes, but also to precisely control their polarization direction. The latter remains challenging because the net spin Hall response of a nonmagnetic source is constrained by macroscopic symmetry [6]. Once the material orientation and charge-current direction are fixed, the symmetry-allowed spin-current polarization is also fixed. This symmetry locking is a central limitation of conventional high-performance spin Hall materials: for example, a conventional in-plane-current geometry typically produces an out-of-plane-flowing spin current with an in-plane spin polarization, whereas many applications, including deterministic spin-orbit torque (SOT) manipulation of perpendicular magnets [7–9], require additional tilted polarization components.

Because of this apparent immutability of spin Hall polarization, existing strategies usually circumvent rather than reprogram the spin source. External magnetic fields [10–12], exchange bias [13,14] or specially designed device geometries [15–21] can provide the missing symmetry breaking, but at the cost of increased device complexity. Alternatively, one may replace conventional spin Hall metals with low-symmetry materials that allow unconventional spin polarizations [22–29]. This route directly addresses the symmetry requirement, but it often sacrifices the large conductivity, robustness, and process compatibility of conventional high-symmetry spin sources [10–12,30].

This raises a fundamental question: can the polarization direction of a spin Hall current be reprogrammed without changing the nonmagnetic bulk spin source itself? Here we show that the answer is yes once the spin Hall current is viewed not as a bulk response alone, but as an emitted spin current jointly determined by the bulk spin Hall texture and the interfacial transmission process. From this perspective, we formulate emitted spin currents as bulk–interface hybrid responses and propose interfacial noncollinear filtering as the mechanism underlying this conversion: a low-symmetry interfacial spin-orbit field imposes spin-dependent transmission on momentum-resolved incident spin Hall channels and selectively transmits hidden spin-polarization components. As a result, conventional high-symmetry spin Hall sources can emit unconventional spin currents forbidden by their bulk spin Hall tensors.

We first revisit the spin Hall effect from the viewpoint of an emitted spin current in a spin-orbit-torque device [6]. In a typical spintronic heterostructure, when a charge current is applied along the $x$ direction, spin currents flowing along the transverse directions can in principle be generated. Among them, the spin current flowing along the $z$ direction is particularly important, because it can cross the interface and enter the adjacent ferromagnetic layer, where it exerts spin-orbit torque on the magnetic order.

Microscopically, a spin current can be understood as the combined motion of electrons with opposite spin polarizations in opposite directions. The conventional bulk spin Hall response is characterized by the spin Hall conductivity tensor $\sigma_{ij}^{k}$, where $j$ denotes the charge-current direction, $i$ denotes the spin-current flow direction, and $k$ denotes the spin-polarization direction. Within the intrinsic Berry-curvature formalism, it can be written as [5]

$$\sigma_{ij}^{k} = e\hbar \sum_{n} \int \frac{d^3\mathbf{k}}{(2\pi)^3} f_{n\mathbf{k}} \Omega_{ij,n}^{k}(\mathbf{k}), \quad (1)$$

where $f_{n\mathbf{k}}$ is the Fermi-Dirac distribution function and $\Omega_{ij,n}^{k}(\mathbf{k})$ is the spin Berry curvature of band $n$,

$$\Omega_{ij,n}^{k}(\mathbf{k}) = -2\, Im \sum_{m \neq n} \frac{\langle \psi_{n\mathbf{k}} | \hat{J}_i^k | \psi_{m\mathbf{k}} \rangle \langle \psi_{m\mathbf{k}} | \hat{v}_j | \psi_{n\mathbf{k}} \rangle}{(E_{n\mathbf{k}} - E_{m\mathbf{k}})^2}. \quad (2)$$

Here $\psi_{n\mathbf{k}}$ and $E_{n\mathbf{k}}$ are the Bloch wave function and energy of band $n$ at momentum $\mathbf{k}$, respectively, $\hat{v}_j$ is the velocity operator along $j$, and $\hat{J}_i^k = \{\hat{v}_i, \hat{s}_k\}/2$ is the spin-current operator for a spin current flowing along $i$ with spin component indexed by $k$. This bulk quantity describes the strength of the spin current generated inside the spin source.

The bulk spin Hall conductivity (Eq. (1)) is a fully momentum-integrated quantity. However, for an individual electronic state, the spin polarization is not required to be along the macroscopic spin Hall polarization direction; it can depend on band index, momentum, and propagation direction. After summing over all symmetry-related momenta in the bulk, these microscopic spin components are constrained by the macroscopic crystal symmetry. In conventional high-symmetry spin Hall materials, for example, when the charge current flows along $x$, the allowed spin Hall response is such that the charge-current direction, spin-current flow direction, and spin-polarization direction are mutually orthogonal. In a standard geometry, this gives the conventional component $\sigma_{zx}^{y}$, corresponding to a $z$-flowing spin current with $y$-spin polarization.

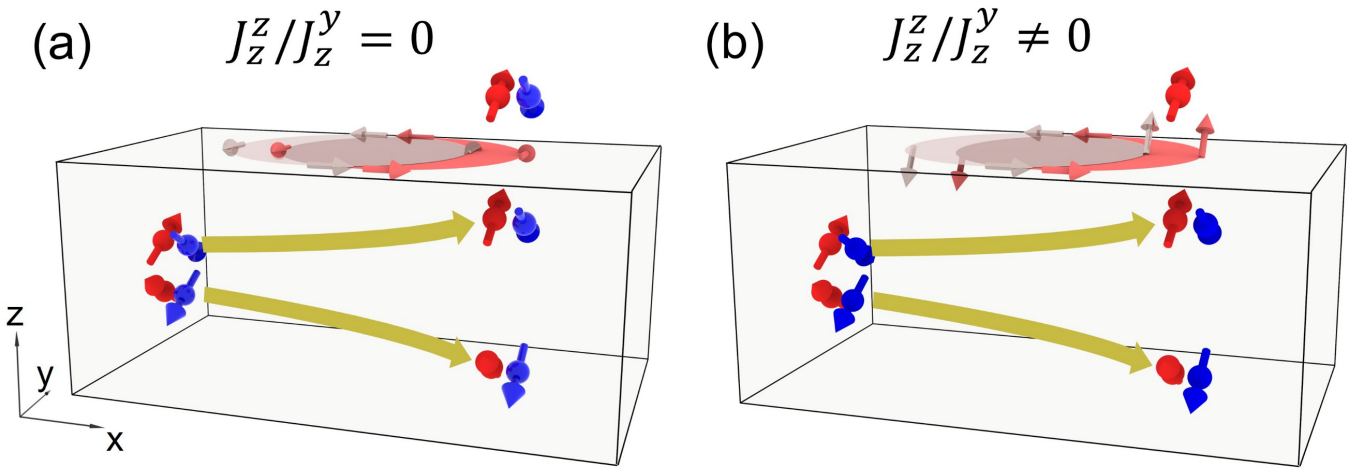

**Fig. 1: Physical picture of interfacial noncollinear filtering of a spin Hall current.** The yellow arrows indicate the flow direction of the bulk spin Hall current, while the large red and blue arrows denote spin Hall channels carrying different spin components. Although these channels may contain different spin directions, the symmetry-averaged bulk spin Hall current has only a net $y$-spin polarization. The top surface represents the emitting interface, where the small arrows on the Fermi surfaces denote the interfacial spin texture; its electric-field-induced nonequilibrium spin accumulation acts as the spin-orbit filtering field. (**a**) At a high-symmetry interface, the filtering field lies in the interface plane and preserves the compensation between hidden opposite out-of-plane spin components, giving $J_z^z = 0$. (**b**) At a low-symmetry interface, the filtering field acquires out-of-plane components and noncollinearly filters the incident spin Hall channels, making the compensation incomplete and generating a finite $J_z^z$.

In a realistic spin-orbit-torque device with a magnetic layer deposited on the spin source layer, the spin current that acts on the magnetic layer is not simply the fully symmetry-dictated bulk spin Hall current. Only the electrons propagating toward the interface can be injected into the adjacent magnetic layer. Therefore, the relevant incident spin current should be defined from those states with positive group velocity along the interface normal. Without loss of generality, for a $z$-normal interface the corresponding momentum-resolved incident spin Hall response is

$$\begin{aligned}\sigma_{zx,n}^{k,>}(\mathbf{k}_\parallel) &= \frac{e\hbar}{2\pi} \int dk_z \sigma_{zx,n}^{k,>}(\mathbf{k}) \\ &= \frac{e\hbar}{2\pi} \int dk_z f_{n\mathbf{k}} \Omega_{zx,n}^{k}(\mathbf{k}) \Theta\big[v_{z,n}(\mathbf{k})\big] \end{aligned} \quad (3)$$

where $v_{z,n}(\mathbf{k})$ is the $z$-component of the group velocity of band $n$, $\Theta$ is the step function, $\mathbf{k}_\parallel = (k_x, k_y)$ is the transverse momentum, and the superscript $>$ denotes the incident states with $v_{z,n} > 0$. We define the spin-polarization vector $\boldsymbol{\sigma}_{zx,n}^{>}(\mathbf{k}_\parallel) = \big(\sigma_{zx,n}^{x,>}(\mathbf{k}_\parallel), \sigma_{zx,n}^{y,>}(\mathbf{k}_\parallel), \sigma_{zx,n}^{z,>}(\mathbf{k}_\parallel)\big)$ and its direction $\hat{\mathbf{p}}_n(\mathbf{k}_\parallel) = \boldsymbol{\sigma}_{zx,n}^{>}(\mathbf{k}_\parallel)/\big|\boldsymbol{\sigma}_{zx,n}^{>}(\mathbf{k}_\parallel)\big|$, which retains the information of the momentum-resolved spin Berry curvature carried by the states moving toward the interface. Components that vanish in the fully symmetry-dictated bulk tensor may therefore still exist in $\boldsymbol{\sigma}_{zx,n}^{>}(\mathbf{k}_\parallel)$.

When these incident electrons cross the interface, they are further acted on by the interfacial spin-orbit field [19,31,32]. This field is generated by the interfacial Edelstein response and is dictated by interfacial symmetry. We characterize it by the momentum-resolved spin Edelstein susceptibility [6,33–35]

$$m_k(\mathbf{k}_\parallel) = \chi_{kx}(\mathbf{k}_\parallel)\mathcal{E}_x, \tag{4}$$

with

$$\chi_{kx}(\mathbf{k}_\parallel) = \frac{A_0 g_s \mu_B e}{4\pi^3} \sum_{n\neq m} \frac{\Gamma^2 Re[\langle\phi_{n\mathbf{k}}|\hat{s}_i|\phi_{m\mathbf{k}}\rangle\langle\phi_{m\mathbf{k}}|\hat{v}_x|\phi_{n\mathbf{k}}\rangle]}{[(\epsilon_F - \epsilon_{n\mathbf{k}})^2 + \Gamma^2][(\epsilon_F - \epsilon_{m\mathbf{k}})^2 + \Gamma^2]}. \tag{5}$$

Here $m_k(\mathbf{k}_\parallel)$ is the $k$-component of the current-induced interfacial spin polarization, the applied $\mathcal{E}_x$ is assumed along the $x$ direction, $A_0$ is the interface unit-cell area, $g_s$ is the spin $g$-factor, $\mu_B$ is the Bohr magneton, $\epsilon_F$ is the Fermi energy, and $\Gamma = 1$ meV is the disorder broadening. The wave functions $\phi_{n\mathbf{k}}$ and energies $\epsilon_{n\mathbf{k}}$ in the Edelstein susceptibility both refer to interfacial electronic states. $\widehat{\mathbf{m}}(\mathbf{k}_\parallel) = \mathbf{m}(\mathbf{k}_\parallel)/|\mathbf{m}(\mathbf{k}_\parallel)|$ defines the direction of the local interfacial spin-orbit field. The effect of the interface on the incident electrons can then be described by a spin-dependent delta-function barrier [19,31],

$$V(\mathbf{k}_\parallel, z) = \frac{\hbar^2 k_F}{m_e}\delta(z)[u_0 - u_R(\mathbf{k}_\parallel)\mathbf{s}\cdot\widehat{\mathbf{m}}(\mathbf{k}_\parallel)], \tag{6}$$

Here $k_F$ is the Fermi wave vector of the incident electron, $m_e$ is the electron mass, $u_0 = 1$ is the spin-independent part of the interface barrier, and $u_R(\mathbf{k}_\parallel)_{max} = 0.1u_0$ is the strength of the spin-dependent interfacial spin-orbit term.

Solving the scattering problem gives the spin-dependent transmission probability of the electron on the band $n$(see Sec. II A of the Supplemental Material) [36]

$$T_n(\mathbf{k}_\parallel) = \frac{k_z^4 + k_z^2 k_F^2 u_0^2\left[1 + \frac{u_R(\mathbf{k}_\parallel)}{u_0}\widehat{\mathbf{p}}_n(\mathbf{k}_\parallel)\cdot\widehat{\mathbf{m}}(\mathbf{k}_\parallel)\right]^2}{\left[k_F^2\left(u_0^2 - u_R^2(\mathbf{k}_\parallel)\right) - k_z^2\right]^2 + 4k_z^2 k_F^2 u_0^2}. \tag{7}$$

Here $k_z$ is the normal component of the incident wave vector. Equation (7) shows that the transmission is governed by the relative orientation between the incident spin polarization vector and the interfacial spin-orbit field, even when the two are noncollinear. We thus refer to this effect as *interfacial noncollinear spin-orbit filtering*. Electrons whose spin polarization is closer to the local interfacial field are transmitted more efficiently, whereas strongly misaligned channels are more strongly suppressed. The spin current emitted into the ferromagnet is therefore not the incident bulk spin Hall current itself, but an emitted spin current from bulk–interface hybrid responses.

For a conventional high-symmetry Rashba interface [37–39], the interfacial spin-orbit field lies in the plane and usually changes only the magnitude of the transmitted spin current [19], leaving $\mathbf{k}_\parallel$-resolved out-of-plane components compensated after momentum integral (Fig. 1(a)). At a low-symmetry interface, however, the Edelstein spin texture can acquire an out-of-plane component [40–42]. The interface can then transmit incident channels with opposite $\mathbf{k}_\parallel$-resolved out-of-plane spin components unequally, making their cancellation incomplete and generating an emitted spin current whose polarization differs from that of the bulk spin Hall current (Fig. 1(b)).

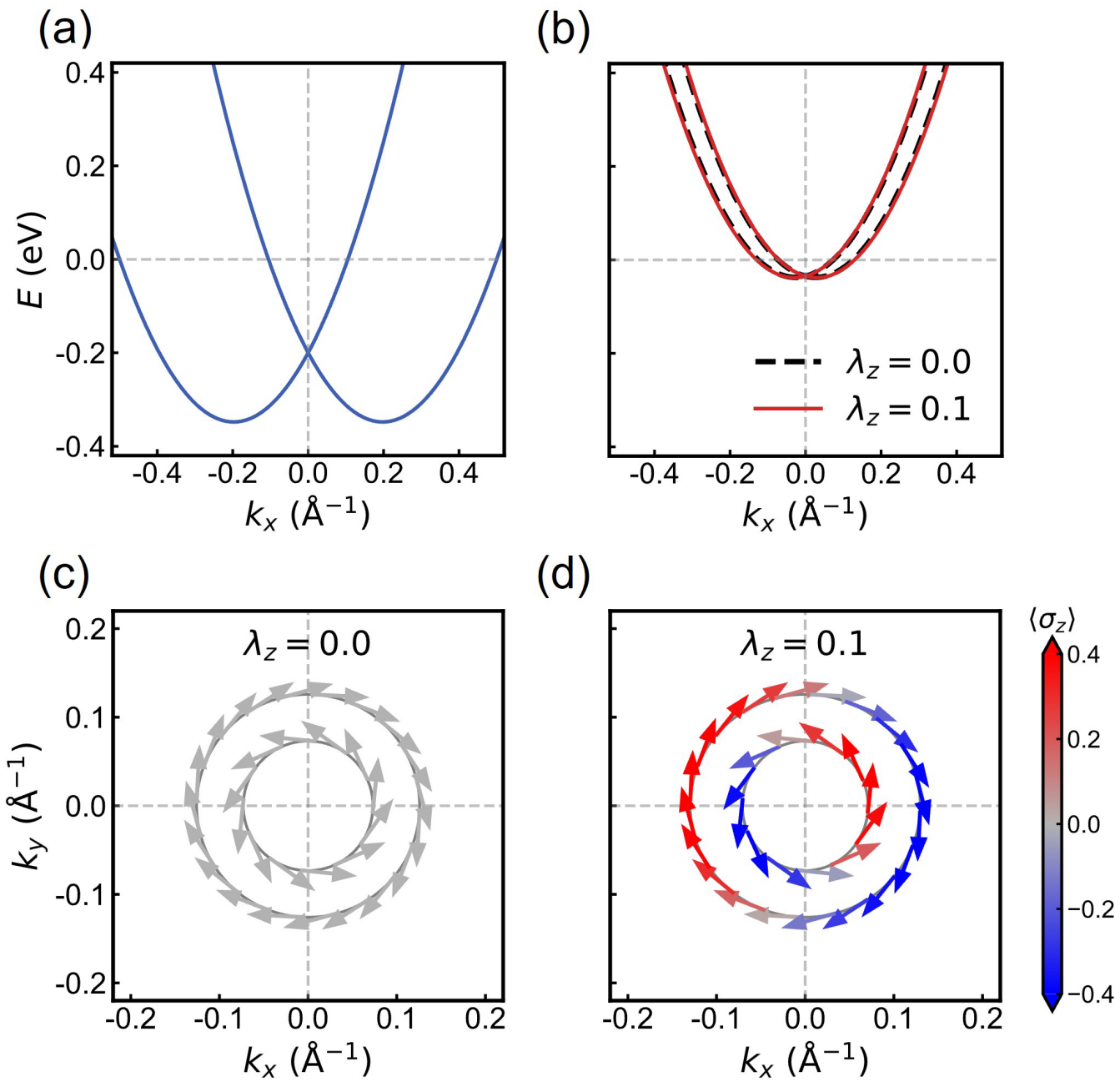

**Fig. 2: Minimal model for interfacial noncollinear filtering.** (**a**) Bulk band structure of the high-symmetry Dirac spin Hall source. (**b**) Interface dispersion for the conventional Rashba case and for the generalized Rashba model with finite $\lambda_z$. (**c**) In-plane spin texture of the conventional Rashba interface. (**d**) Noncollinear interfacial spin texture generated by finite $\lambda_z$.

Finally, the emitted spin current is estimated by weighting the $\boldsymbol{\sigma}^{>}_{zx,n}(\mathbf{k}_\parallel)$ with the spin-dependent transmission probability. We define

$$G_z^k(\mathbf{k}_\parallel) = \sum_n \left[\left|\boldsymbol{\sigma}^{>}_{zx,n}(\mathbf{k}_\parallel)\right| T_n(\mathbf{k}_\parallel)\, p_{n,k}(\mathbf{k}_\parallel)\right]_{E_F}, \tag{8}$$

where $p_{n,k}$ is the $k$-component of $\widehat{\mathbf{p}}_n$. The normalized emitted spin current is then

$$J_z^k = \frac{eV}{\mathcal{N}}\int d^2\mathbf{k}_\parallel G_z^k(\mathbf{k}_\parallel). \tag{9}$$

Here $V$ is the applied bias, and $\mathcal{N} = 4\pi\int d^2\mathbf{k}\sum_n\left|\boldsymbol{\sigma}^{>}_{zx,n}(\mathbf{k})\right|_{E_F}$ is a normalization factor determined by the total magnitude of the incident spin Hall response. For comparing the polarization of the emitted spin current, we define the ratio $J_z^z/J_z^y$.

Eqs. (8,9) make explicit that the emitted spin current is controlled by three ingredients: the momentum-resolved spin Berry curvature of the bulk, the velocity selection associated with propagation toward the interface, and the spin-dependent transmission imposed by the interfacial spin-orbit texture. The interface does not generate the bulk spin Hall current; it converts the $\mathbf{k}_\parallel$-resolved spin Hall texture into a transmitted spin current with an unexpected polarization.

To demonstrate the above mechanism, we first consider a minimal model consisting of a high-symmetry bulk spin Hall source and a low-symmetry spin-orbit-coupled interface. The bulk is described by an isotropic Dirac Hamiltonian [43],

$$H(\mathbf{k})_B = v\tau_x \mathbf{k}\cdot\boldsymbol{\sigma} + C\mathbf{k}^2 - V_B, \tag{10}$$

where $\sigma$ and $\tau$ are Pauli matrices acting on spin and orbital degrees of freedom, respectively. The parameters $v = 1.5$ eVÅ, and $C = \hbar^2/2m$ denote the Dirac velocity and quadratic correction, respectively, while $V_B = 0.2$eV is a spin-independent potential. The bulk band structure of Eq. (10) is shown in Fig. 2(a). Owing to its continuous rotational symmetry, this bulk Hamiltonian does not allow the unconventional components in a macroscopic spin Hall tensor.

The interface is modeled by a generalized Rashba Hamiltonian [44],

$$H(\mathbf{k})_I = \lambda(k_x\sigma_y - k_y\sigma_x) + \lambda_z k_x \sigma_z + C\mathbf{k}_\parallel^2 - V_I. \tag{11}$$

The coefficient $V_I = 0.035$ eV is the interfacial potential offset, $\lambda = 0.2$ eVÅ is the conventional in-plane Rashba coupling, and $\lambda_z$ characterizes the out-of-plane component of the interfacial spin-orbit texture. The band dispersion and spin texture given by Eq. (11) are shown in Fig. 2(b-d). When $\lambda_z = 0$, Eq. (11) reduces to the conventional Rashba model, leading to the interfacial spin texture that lies entirely in the plane (Fig. 2(c)). In particular, the term $\lambda_z k_x \sigma_z$ with $\lambda_z \neq 0$ is allowed by a reduced interfacial symmetry with a single mirror plane, and the resulting spin-orbit texture can have the out-of-plane component (Fig. 2(d)).

We next combine the bulk Hamiltonian in Eq. (10) and the interfacial Hamiltonian in Eq. (11) to construct a minimal spin-orbit-torque device geometry, as illustrated in Fig. 3(a). The bulk layer acts as the spin Hall source, while the interface selects and filters the spin current propagating toward the adjacent ferromagnetic layer.

We first examine the $\mathbf{k}_\parallel$-resolved spin Hall response generated by the bulk Hamiltonian. Because the bulk model has continuous rotational symmetry, the $\mathbf{k}_\parallel$-resolved $x$-polarized response vanishes, $\sigma_{zx,n}^{x,>}(\mathbf{k}) = 0$ (see Sec. III A of the Supplemental Material) [36]. The conventional $y$-polarized component, $\sigma_{zx,n}^{y,>}(\mathbf{k}_\parallel)$, has a distribution that is symmetric with respect to the zone center. Although two bands crossing the Fermi level contribute with opposite signs over the relevant momentum region, as shown in Figs. 3(b,c), their contributions do not completely cancel after summation. Consequently, the incident conventional spin Hall response remains finite, $\sigma_{zx}^{y,>} \neq 0$. By contrast, the out-of-plane component satisfies $\sigma_{zx,n}^{z,>}(k_x, k_y) = -\sigma_{zx,n}^{z,>}(k_x, -k_y)$, as shown in Figs. 3(f,g). The momentum-space cancellation enforced by this antisymmetry leads to a vanishing net out-of-plane incident response, $\sigma_{zx}^{z,>} = 0$, consistent with the macroscopic symmetry constraint of the bulk spin Hall tensor.

The interface determines whether this $\mathbf{k}_\parallel$-resolved out-of-plane component can survive transmission. For a conventional Rashba interface with $\lambda_z = 0$, the interfacial spin-orbit filtering texture is confined to the plane (see Sec. II C of the Supplemental Material)[36]. In this case, the out-of-plane Edelstein component $\chi_{zx}(\mathbf{k}_\parallel)$ is absent, and the relevant in-plane component $\chi_{yx}(\mathbf{k}_\parallel)$ obeys the mirror-related symmetry $\chi_{yx}(k_x, k_y) = \chi_{yx}(-k_x, k_y)$. Therefore, matching the $\mathbf{k}_\parallel$-resolved incident spin Hall response $\sigma_{zx,n}^{k,>}(\mathbf{k}_\parallel)$ with the interfacial filtering texture $\chi_{kx}(\mathbf{k}_\parallel)$ does not break the compensation of the $\mathbf{k}_\parallel$-resolved $z$-spin component. The transmitted conventional component $G_z^y(\mathbf{k}_\parallel)$ can remain finite after momentum integration, whereas the transmitted out-of-plane component keeps an antisymmetric distribution, $G_z^z(k_x, k_y) = -G_z^z(k_x, -k_y)$, and thus gives $J_z^z = 0$ (see Sec. II C of the Supplemental Material) [36].

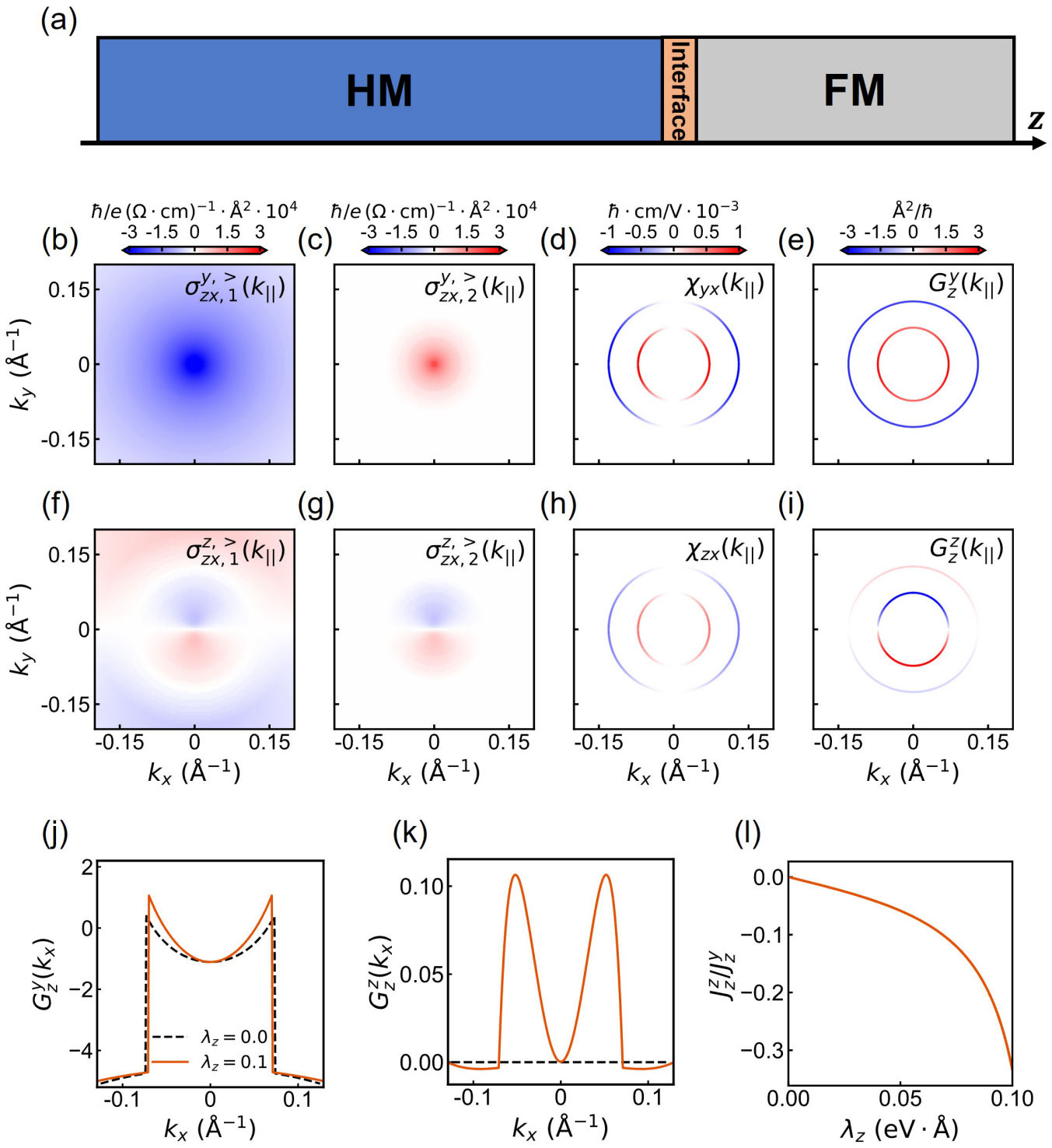

**Fig. 3: Momentum-resolved generation of an unconventional emitted spin current by interfacial noncollinear filtering.** (**a**) The schematic of a SOT device supporting interfacial noncollinear filtering. (**b,c**) $\mathbf{k}_\parallel$-resolved conventional spin Hall conductivity $\sigma_{zx,n}^{y,>}(\mathbf{k}_\parallel)$ for $n$=1 (**b**) and $n$=2 (**c**). (**d**) In-plane component $\chi_{yx}(\mathbf{k}_\parallel)$ of the interfacial spin-orbit filtering texture. (**e**) Momentum resolved contribution $G_z^y(\mathbf{k}_\parallel)$ to the emitted conventional spin current. (**f,g**) $\mathbf{k}_\parallel$-resolved unconventional spin Hall conductivity $\sigma_{zx,n}^{z,>}(\mathbf{k}_\parallel)$ for $n$=1 (**f**) and $n$=2 (**g**). (**h**) Out-of-plane component $\chi_{zx}(\mathbf{k}_\parallel)$ of the interfacial spin-orbit filtering texture. (**i**) Momentum resolved contribution $G_z^z(\mathbf{k}_\parallel)$ to the emitted out-of-plane spin current. (**j,k**) $k_y$-integrated transmitted contributions $G_z^y(k_x)$ (**j**) and $G_z^z(k_x)$ (**k**). (**l**) $J_z^z/J_z^y$ as a function of $\lambda_z$. Here, $\lambda_z = 0.1$ eVÅ is used in (d,e,h,i).

For finite out-of-plane interfacial spin-orbit coupling, ($\lambda_z = 0.1$ ), the interface develops a finite out-of-plane Edelstein component $\chi_{zx}(\mathbf{k}_\parallel)$ (Fig. 3(h)), while the in-plane component $\chi_{yx}(\mathbf{k}_\parallel)$ remains finite (Fig. 3(d)). The filtering field is then noncollinear with the conventional incident spin polarization and gives different transmission probabilities to channels carrying opposite $\mathbf{k}_\parallel$ -resolved $z$-spin components. This breaks the antisymmetric compensation of $G_z^z(\mathbf{k}_\parallel)$ , producing a finite transmitted out-of-plane contribution (Fig. 3(i)), while the conventional contribution $G_z^y(\mathbf{k}_\parallel)$ is only weakly modified (Fig. 3(e)). The distributions summed over the $k_y$ direction, $G_z^s(k_x) = \sum_{k_y} G_z^s(\mathbf{k}_\parallel)$ further show that $G_z^y(k_x)$ remains sizable for both $\lambda_z = 0$ and 0.1 eVÅ, whereas $G_z^z(k_x)$ becomes finite only when $\lambda_z \neq 0$ (Figs. 3(j,k)). Consequently, the emitted spin-current ratio $J_z^z/J_z^y$ increases monotonically with $\lambda_z$ and reaches about 0.3 at $\lambda_z = 0.1$ eVÅ (Fig. 3(l)). This confirms that an emitted spin current with a strongly tilted spin polarization can be generated even when the macroscopic bulk spin Hall tensor forbids unconventional components. Moreover, as shown in the end matter, the $J_z^z/J_z^y$ ratio can be further enhanced by choosing spin source materials with relatively larger $\Omega_{n,zx}^{z,>}(\mathbf{k})$, even with a small $\lambda_z$.

Our theory incorporates two effects that naturally coexist in spin-orbit-torque heterostructures: the bulk spin Hall effect in the spin source and the interfacial spin-orbit coupling at the boundary. Treating the emitted spin current as a bulk-interface hybrid response shows that the final spin polarization injected into the magnetic layer is not uniquely determined by the bulk spin Hall tensor. Forbidden components in the symmetry-dictated bulk response can survive at the $\mathbf{k}_\parallel$-resolved level and be selected by spin-dependent interfacial transmission. This mechanism is also distinct from conventional spin filtering [19,45,46], where the outgoing polarization is usually dictated by the effective field or magnetization of the filter. Here the interface does not impose a fixed spin direction; it provides asymmetric momentum-dependent transmission, while the bulk supplies the $\mathbf{k}_\parallel$-resolved spin Berry-curvature texture.

More broadly, our work suggests a way to formulate responses in composite quantum systems. In many heterostructures, different components are often treated separately, or the effect of one component is incorporated into another through an effective self-energy or non-Hermitian description. Here we instead focus on the emitted current as an intrinsically bulk–interface hybrid response: it is neither a property of the bulk spin Hall tensor alone nor a response imposed solely by the interface, but a transport quantity jointly shaped by the bulk spin Hall texture and the interfacial transmission process. This viewpoint may extend beyond spin currents to other angular-momentum currents, such as valley and orbital Hall currents [47–50].

This viewpoint also suggests a practical route for experiments. Instead of replacing conventional spin Hall metals with intrinsically low-symmetry materials, one can use efficient high-symmetry spin sources such as Pt, Ta, W, and related heavy metals, and engineer their interfacial symmetry through growth orientation [51], vicinal surfaces, stepped interfaces, or symmetry-breaking interlayers. Because the minimal model shows that the bulk source itself can even have continuous rotational symmetry, this strategy may also apply to polycrystalline or amorphous spin sources capped by a low-symmetry two-dimensional (2D) material. Especially, the 2D materials with Ising-type spin-orbit coupling ($\lambda = 0$ and $\lambda_z \neq 0$) [52–54] may allow ultra-strong interfacial filtering(see Sec. II E and II F of the Supplemental Material) [36] . Using ferroelectric 2D materials as the capping layer further offers the electric-tunable interfacial filtering, since the spin-orbit textures of these materials can be reversed by ferroelectric switching [55–57].

Overall, our work challenges the conventional view that the polarization of a spin Hall current is fixed solely by the macroscopic symmetry of the bulk spin source. In realistic heterostructures, the spin current relevant for device operation is an emitted current shaped jointly by the momentum-resolved bulk spin Hall texture and the spin-dependent transmission at the interface. Thus, Interfacial noncollinear filtering provides a mechanism to convert spin-polarization components forbidden by the symmetry-constrained bulk response into observable emitted spin currents, turning conventional high-symmetry spin sources into generators of unconventional spin polarizations. This bulk-interface perspective shifts the control of spin-current polarization from material symmetry alone to interfacial quantum selection, offering a route to realize out-of-plane or tilted spin currents in efficient, CMOS-compatible spin Hall materials without relying on intrinsically low-symmetry bulk crystals or complex external symmetry-breaking schemes.

**Acknowledgments.** We thank Xuepeng Qiu and Yumeng Yang for helpful discussions. This work was supported by the National Key R&D Program of China (Grant Nos. 2024YFB3614100 and 2024YFA1410100), the National Natural Science Foundation of China (Grants Nos. 12241405, 12274411, 12404185, and 12474121), the Basic Research Program of the Chinese Academy of Sciences Based on Major Scientific Infrastructures (Grant No. JZHKYPT-2021-08), and the CAS Project for Young Scientists in Basic Research (Grant No. YSBR-084). Calculations were performed at Hefei Advanced Computing Center.

**Data availability.** The data that support the findings of this article are not publicly available. The data are available from the authors upon reasonable request.

## End matter

Here we test the interfacial noncollinear filtering mechanism using a realistic high-symmetry material as the spin source. We consider the Dirac semimetal $Cd_3As_2$, which provides a representative spin-orbit-coupled system with strong band inversion and high bulk symmetry. The low-energy Hamiltonian is written as [58–60]

$$H_{\mathrm{Cd_3As_2}}(\mathbf{k}) = a(\mathbf{k})\sigma_z s_0 + b(\mathbf{k})\sigma_x s_z + c(\mathbf{k})\sigma_y s_0 + d(\mathbf{k})\sigma_x s_x + e(\mathbf{k})\sigma_x s_y + V_0\sigma_0 s_0, \tag{12}$$

where $s_i$ and $\sigma_i$ are Pauli matrices acting on spin and orbital degrees of freedom, respectively. The coefficients are

$$\begin{aligned} a(\mathbf{k}) &= m_0 - m_1 k_z^2 - m_2(k_x^2 + k_y^2), \\ b(\mathbf{k}) &= \eta k_x, \qquad c(\mathbf{k}) = -\eta k_y, \\ d(\mathbf{k}) &= (\beta+\gamma)k_z(k_y^2 - k_x^2), \\ e(\mathbf{k}) &= -2(\beta-\gamma)k_z k_x k_y. \end{aligned} \tag{13}$$

This effective Hamiltonian respects the tetragonal symmetry and contains the mirror operations $M_x$, $M_y$, and $M_z$. These symmetries forbid a fully symmetry-averaged unconventional bulk spin Hall response, even though the $\mathbf{k}_\parallel$-resolved response can contain sizable hidden components.

The calculated low-energy band structure is shown in Fig. 4(a). Dirac points appear along the $\Gamma - Z$ direction, and the dispersion near the Dirac points can be tuned by the mass parameter $m_1$. Increasing $m_1$ changes the slope of the Dirac bands near the Fermi level, thereby modifying the distribution of the $\mathbf{k}_\parallel$-resolved spin Berry curvature that enters the incident spin Hall response.

Figure 4(b) shows the emitted spin-current ratio $J_z^z/J_z^y$ as a function of the out-of-plane interfacial spin-orbit coupling $\lambda_z$. Similar to the minimal model, the ratio increases nearly linearly with $\lambda_z$. This indicates that even a relatively small out-of-plane component of the interfacial spin-orbit field can generate a sizable out-of-plane emitted spin current once the incident spin Hall texture provides a favorable hidden $z$-spin component. Moreover, for the same $\lambda_z$, the ratio becomes larger when $m_1$ is increased.

The microscopic origin of this enhancement is revealed by the $\mathbf{k}_\parallel$-resolved spin Hall conductivities shown in Figs. 4(c) and 4(d). Near the Dirac points, the conventional component $\sigma_{zx}^{y,>}(\mathbf{k}_\parallel)$ is relatively weak, whereas the hidden out-of-plane component $\sigma_{zx}^{z,>}(\mathbf{k}_\parallel)$ is much larger over the relevant momentum region. Although the net unconventional bulk response is still forbidden by the mirror symmetries, this large $\mathbf{k}_\parallel$-resolved hidden component provides an efficient source for interfacial filtering. Once the low-symmetry interface supplies a finite $\chi_{zx}(\mathbf{k}_\parallel)$ and hence a finite $\lambda_z$, the hidden $\sigma_{zx}^{z,>}(\mathbf{k}_\parallel)$ texture can be converted into a large emitted $J_z^z$.

These results provide a practical design principle for generating unconventional emitted spin currents from high-symmetry spin Hall sources. A material does not need to support a macroscopic unconventional spin Hall tensor in the bulk. Instead, it can serve as an efficient unconventional emitted spin-current source if it hosts a large momentum-resolved hidden spin Berry-curvature component, such as $\Omega_{n,zx}^{z,>}(\mathbf{k})$, that can be selected by a low-symmetry interfacial spin-orbit filter. Thus, the key requirement is the combination of a favorable hidden bulk spin Berry-curvature texture and a symmetry-matched interfacial noncollinear filtering field.

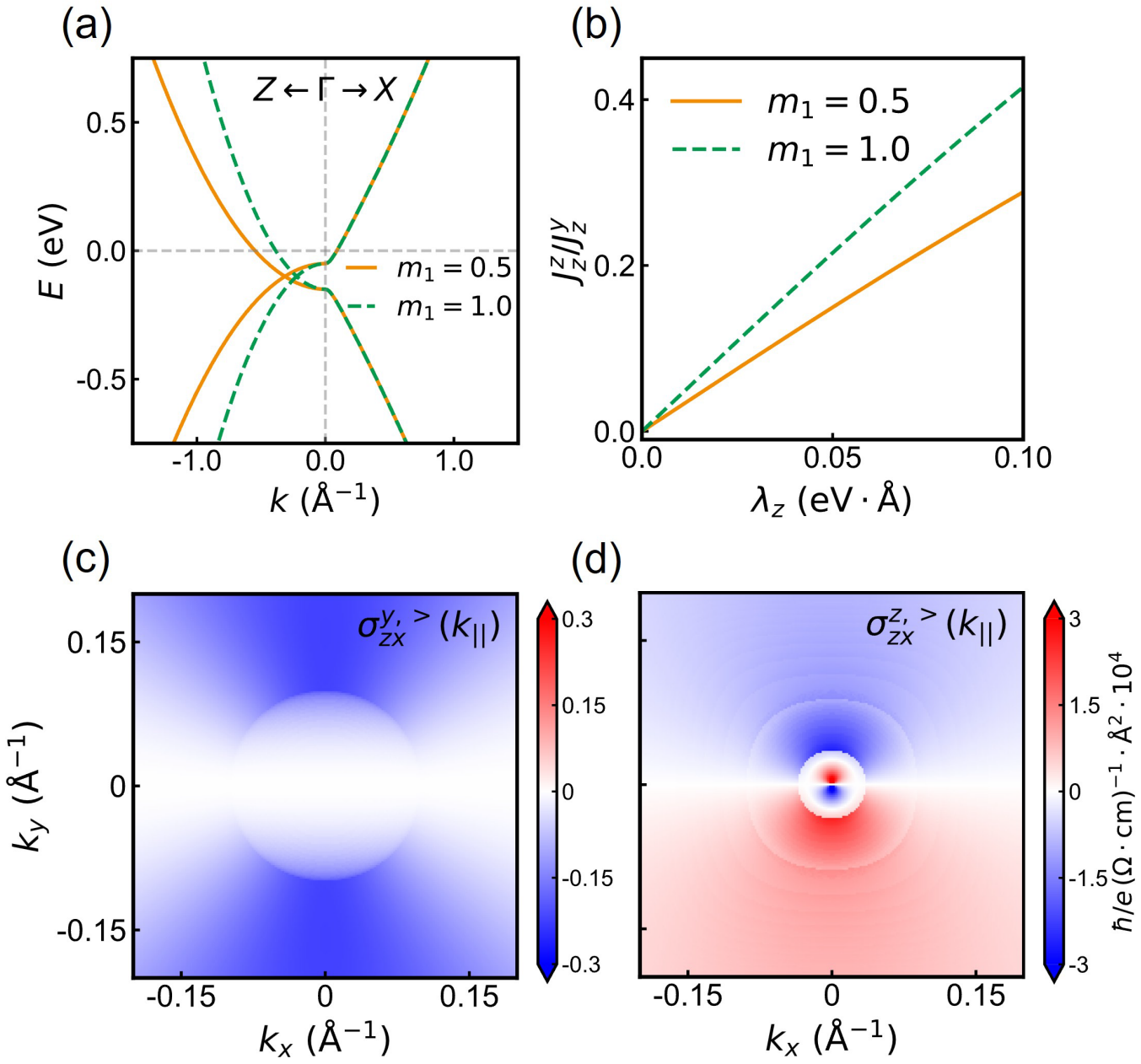

**Fig. 4: Interfacial noncollinear filtering based on the $Cd_3As_2$ Hamiltonian.** (**a**) Low-energy band structure of $Cd_3As_2$ for representative values of the mass parameter $m_1$. Dirac points appear along the $\Gamma$–$Z$ direction, and the dispersion near the Dirac points is tuned by $m_1$. (**b**) Emitted spin-current ratio $J_z^z/J_z^y$ as a function of the out-of-plane interfacial spin-orbit coupling $\lambda_z$. The ratio increases nearly linearly with $\lambda_z$ and is enhanced for larger $m_1$. (**c,d**) $\mathbf{k}_\parallel$-resolved spin Hall conductivities $\sigma_{zx,n}^{y,>}(\mathbf{k}_\parallel)$ and $\sigma_{zx,n}^{z,>}(\mathbf{k}_\parallel)$. The hidden out-of-plane component is much larger than the conventional component near the Dirac points, providing an efficient $\mathbf{k}_\parallel$-resolved spin Berry-curvature texture for interfacial noncollinear filtering.